\documentclass[aps,prl,twocolumn,superscriptaddress]{revtex4-2}

\usepackage{graphicx}
\usepackage{gensymb}
\usepackage{ulem}

\usepackage{color}

\usepackage[colorlinks=true,linkcolor=blue,citecolor=cyan,urlcolor=blue]{hyperref}%

\usepackage{xr}

\begin{document}


\title{Effective bands and band-like electron transport in amorphous solids}



\author{Matthew Jankousky}
\affiliation{Department of Metallurgical and Materials Engineering, Colorado School of Mines, 1500 Illinois St., Golden CO 80401, USA}

\author{Dimitar Pashov}
\affiliation{Theory and Simulation of Condensed Matter, King's College London, The Strand, London WC2R 2LS, UK}

\author{Ross E. Larsen}
\affiliation{Computational Science Center, National Renewable Energy Laboratory, 15013 Denver West Parkway, Golden CO 80401, USA}
\affiliation{Renewable and Sustainable Energy Institute, University of Colorado Boulder, Boulder, Colorado 80309, USA}

\author{Vladimir Dobrosavljevi\'c}
\affiliation{Department of Physics and national High Magnetic Field Laboratory, Florida State University, Tallahassee, FL 32306, USA}

\author{Mark van Schilfgaarde}
\affiliation{Materials, Chemical and Computational Science Directorate, National Renewable Energy Laboratory, 15013 Denver West Parkway, Golden CO 80401, USA}

\author{Vladan Stevanovi{\'c}}
\email[]{vstevano@mines.edu}
\affiliation{Department of Metallurgical and Materials Engineering, Colorado School of Mines, 1500 Illinois St., Golden CO 80401, USA}
\affiliation{Materials, Chemical and Computational Science Directorate, National Renewable Energy Laboratory, 15013 Denver West Parkway, Golden CO 80401, USA}

\date{\today}

\begin{abstract}
The localization of electrons caused by atomic disorder is a well-known phenomenon. However, what circumstances allow electrons to remain delocalized and retain band-like characteristics even when the crystal structure is completely absent, as found in certain amorphous solids, is less well understood. To probe this phenomenon, we developed a fully first-principles description of the electronic structure and charge transport in amorphous solids by combining a novel representation of the amorphous state with the state-of-the-art many-body (QS\textit{GW}) electronic structure theory. Using amorphous In$_2$O$_3$ as an example, we demonstrate the accuracy of our approach in reproducing the band-like nature of the conduction electrons as well as their disorder-limited mobility. Our approach reveals the physical origins responsible for the electron delocalization and the survival of the band dispersions despite the absence of long-range order. 
\end{abstract}

\maketitle


In free space, eigenstates of the Schr\"odinger equation are free particles with well defined momentum $\hbar\mathbf{k}$ that is the constant of motion.  For an electron moving in a perfectly periodic potential, scattering is perfectly coherent. $\mathbf{k}$ remains a good quantum number and, by Bloch's theorem, electron states remain delocalized with $\hbar\mathbf{k}$ reinterpreted as the ``crystal momentum.'' If the potential deviates slightly from periodicity, the Hamiltonian can be partitioned into a large periodic part and a weak remainder.  In such a case transport can be characterized in terms of electrons propagating in states of quantum number $\mathbf{k}$, with scattering between them which can be treated in perturbation theory.  This is the usual description of transport in crystalline systems.  But if the disorder is strong the remainder is not weak and the Bloch description breaks down, as Anderson described in his seminal work~\cite{Anderson_PR:1958}.  

There is another less well-understood possibility that electron states retain their itinerant character even for systems without any translational invariance, which this work tries to address.  Amorphous systems are the archetypal example of broken symmetry and strong disorder. They have no translational symmetry and typically exhibit poor mobility.   But there are exceptions: some amorphous systems have relatively high electron mobility, for example amorphous In$_2$O$_3$ \cite{Bellingham_TSF:1991}.  This system is the base material for a variety of transparent conducting oxides (TCOs) used in a wide range of optoelectronic applications, from photovoltaics to consumer electronics (flat panel displays, channels in thin film transistor structures, etc.) \cite{Ginley_TCOs}. Amorphous TCOs are appealing as they allow for low temperature processing while remaining homogeneous and without grain boundaries. Surprisingly, their electronic properties are comparable to, and sometimes even better than, their (poly)crystalline counterparts. Measured electron mobilities of $\sim$10-40 cm$^2$/Vs or even higher \cite{Ginley_TCOs, Buchholz_CM:2014, Charnas_AM:2024} rival those of polycrystalline films and clearly indicate band-like electron transport in systems without any long-range order. The electron mean free paths of 2-6 nm extracted from resistivity measurements in a-In$_2$O$_3$ \cite{Bellingham_TSF:1991}, present another clear indication of the delocalized nature of electrons in this system. Although electron delocalization and the existence of band-like features in amorphous solids and liquids have previously been discussed, all proposed explanations are largely qualitative and/or phenomenological \cite{Stratt_PRL:1989,Medvedeva_JAP:2020,Lee_PRB:2022}. The complete first-principles description allowing for evaluation of the absolute values of disorder-limited electron mobility in amorphous solids is, to the best of our knowledge, absent from literature.

 \begin{figure*}[t!]
\includegraphics[width=\linewidth]{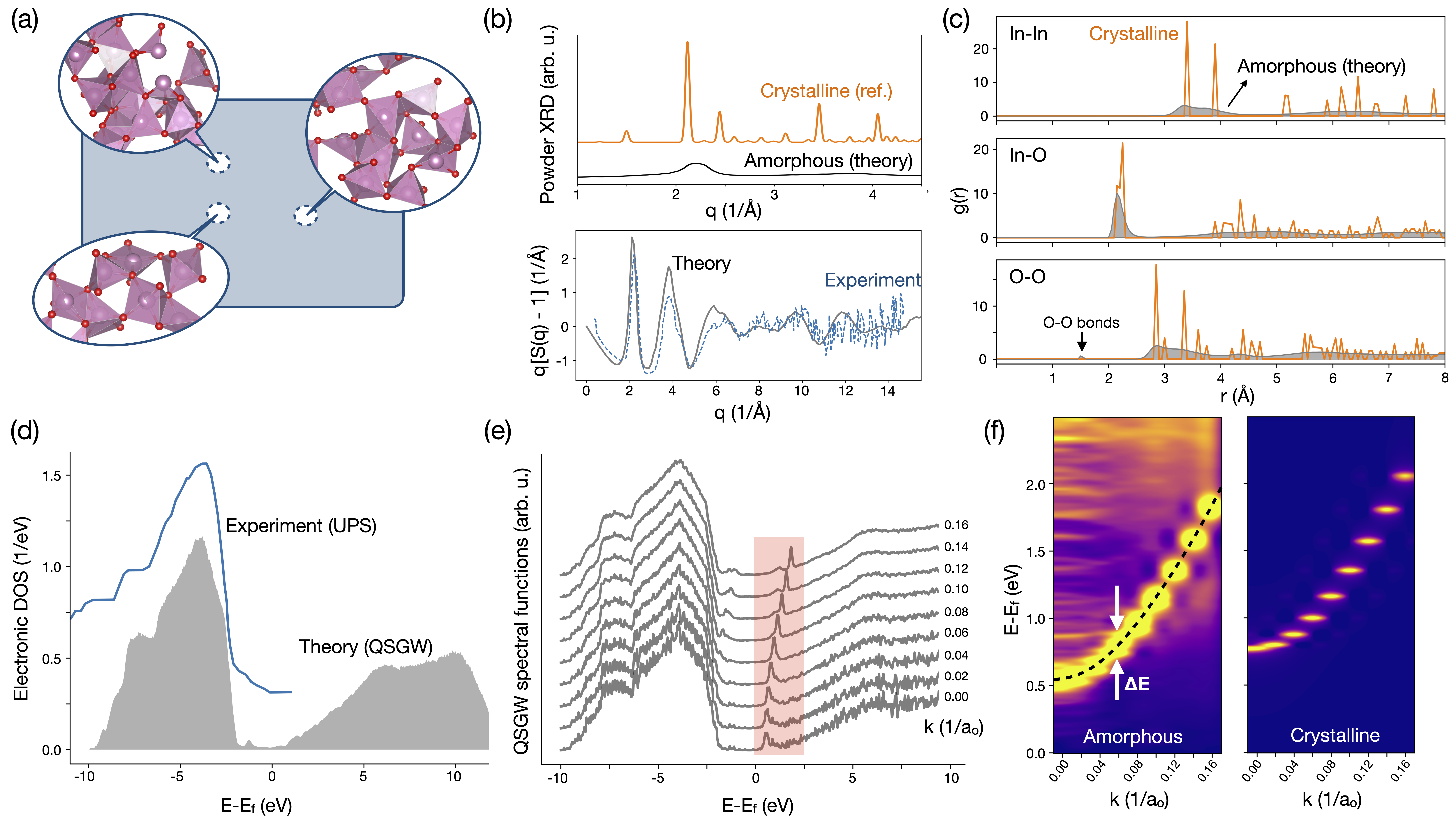}
\caption{\label{fig1} (a) In our work the structure of amorphous solids is represented as a composite of local environments. (b) Calculated XRD intensity and structure factor for a-In$_2$O$_3$ averaged over local environments agree well with the available experimental data. (c) Comparison of the calculated partial pair distribution functions $g(r)$ between crystalline and amorphous In$_2$O$_3$. (d) Calculated (QS\textit{GW}) electronic density of states is shown against photoemission (UPS) measurements. (e) Calculated (QS\textit{GW}) spectral functions as a function of electron energy at different k-points. (f) Spectral functions of the conduction band (red rectangle in (e)) shown as a heat map; the dispersion E(k) is clearly visible and the energy broadening is labeled. A heat map of the spherically averaged conduction band of crystalline In$_2$O$_3$ is also shown for comparison.
}
\end{figure*} 

In this paper we develop a quantitative, fully first-principles description of the effective band structure and the disorder-limited charge-carrier transport in amorphous solids. The approach employs the novel representation of the amorphous state as a composite of local environments \cite{Wolf_JAP:2025}, each corresponding to the small-cell, periodic local minimum on the potential energy surface obtained through the first-principles random structure sampling \cite{Stevanovic_PRL:2016}. As we show here, by averaging over the properties of individual local environments, the measured structural features of a-In$_2$O$_3$ are well reproduced, including the structure factor and the local coordination of atoms. More importantly, this representation enables the use of state of the art many-body electronic structure methods, such as the quasi-particle self-consistent \textit{GW} (QS\textit{GW}) \cite{qsgw_PRL:2004}, which, in turn, allows formulation of an accurate effective (average) band structure for amorphous systems, their electronic density of states, and the direct evaluation of disorder-limited charge carrier mobility from the broadening of the electronic band(s). %
Fig.~\ref{fig1} illustrates our approach and key findings. 

\section{Amorphous solid as a composite of local environments}
%
Here, the structure of the amorphous state is modeled via the first-principles random structure sampling that includes: (i) generation of a large number of structures with a fixed number of atoms, random lattice parameters and random atom positions, and (ii) their relaxation to the closest local minimum on the potential energy surface using density-functional methods. Once relaxed, relevant properties of individual structures are evaluated and the corresponding properties of the amorphous state are obtained via ensemble averaging.   For a-In$_2$O$_3$ presented here, a set of 1,500 random structures with 40 atoms are generated and relaxed using the PBE functional~\cite{GGA}. The number of structures and the cell size were chosen large enough so that the statistical sampling was essentially independent of the choice. It is important to note that the vast majority of relaxed random structures do not exhibit long-range symmetry as evidenced by the space group resolved thermodynamic density of states shown in the supplementary information (SI) Fig.~\ref{TDOS}. 

Fig.~\ref{fig1}(b) depicts structural features of a-In$_2$O$_3$, including both powder XRD intensity and the structure factor $S(q)$, modeled with our approach. The broad feature in the XRD of a-In$_2$O$_3$ located between $q=2-2.5$ {\AA}$^{-1}$ corresponds well to experimental results measured on thin films \cite{Buchholz_CM:2014}. The same is true for the structure factor, as illustrated in Fig.~\ref{fig1}(b), which matches the X-ray diffraction data from Ref.~\cite{Utsuno2006}. All peaks and their intensities for the range of scattering vectors $q$ with relatively low experimental noise are reproduced with satisfactory accuracy. Finally, the local coordination of atoms calculated from the pair distribution functions (PDF) $g(r)$ shown in Fig.~\ref{fig1}(c) agree well with those reported in the literature from both experiments \cite{Buchholz_CM:2014,Utsuno2006,Eguchi2010} and molecular dynamics simulations \cite{Medvedeva_JAP:2020,Aliano2011}. The general shape of the partial PDFs follows that of the crystalline In$_2$O$_3$ (bixbyite structure) with some differences in the coordination numbers.

In each random structure, In atoms have on average between 4.4 and 5.8 O atoms in the first coordination shell, with an ensemble average of  5.3. This indicates that the majority of In atoms are either 5- or 6-fold coordinated by O. This is  smaller than the crystalline bixbyite phase of In$_2$O$_3$, for which each In has 6 O nearest neighbors, and it is consistent with experiments and MD simulations \cite{Buchholz_CM:2014,Utsuno2006,Eguchi2010,Medvedeva_JAP:2020,Aliano2011}. The $2^\mathrm{nd}$ neighbor In-In shell has an average coordination number of 11.5, with the averages of each individual structure ranging between 9.1 and 12.7 --- again, smaller than the twelvefold coordination of bixbyite.  Of the 12 In-In neighbors in bixbyite 6 are separated by 3.34\,{\AA} (edge-sharing In-O polyhedra), while the other 6 are separated by 3.82\,{\AA} (corner-sharing In-O polyhedra).  In the random structure there is a bimodal distribution with the two maxima near 3.4\,{\AA} and 3.7\,{\AA} respectively. In the random structures, the O atoms have predominantly three or four In nearest neighbors with an average first-shell coordination of 3.6. All this implies a strong tendency  of amorphous In$_2$O$_3$ toward In-O coordination, formation of 5- and 6-fold In coordination polyhedra that are connected in the corner- and edge-sharing pattern similar to the crystalline phase. An important feature of the amorphous phase, distinct from the crystalline, is the presence of a small O-O peak at $\sim$1.5 {\AA}, indicating the presence of O-O bonds as also observed in previous studies \cite{Medvedeva_JAP:2020,Aliano2011}.
%
\section{Electronic structure and transport properties in the amorphous phase}
%
Within our approach, calculations of the electronic structure of the amorphous phase entail calculations of the electronic structure of each individual 40-atom structure and subsequent ensemble averaging, making the structural disorder quenched from the point of view of electronic structure.\cite{Stratt_PRL:1989} In this way, spatial fluctuations in the electron energies are naturally included without the need for large supercells. However, while using 40-atom structures allows for the high-accuracy, but computationally demanding QS\textit{GW} calculations, doing it for the entire set of 1,500 random structures is costly.  Hence, the QS\textit{GW} calculations are performed on a randomly chosen subset of 100 structures. As shown in SI Figs.~\ref{sq_conv} and \ref{DOS_conv}, the subset of 100 structures used in QS\textit{GW} electronic structure calculations reproduces well the key structural characteristics of the entire set as well as the electronic density of states at the DFT level of theory. This reduction does lead to an increased level of noise present in ensemble averages, which, while noticeable, is of relatively small amplitude.

Another important aspect of our approach is the alignment of the electronic spectra of different random structures. The picture of the amorphous phase as a composite of local environments suggests electrical contact between the environments and consequently the equilibration of the Fermi energy (E$_f$) across the entire sample. Because of this, before ensemble averaging, all electronic structures are aligned to the same E$_f$ value, which is obtained from the calculated electronic DOS and the charge neutrality condition at low temperature. The validity of this approach is evident from Fig.~\ref{fig1}(d) showing overall good correspondence of the QS\textit{GW} calculated, ensemble average electronic DOS and photoemission (UPS) experiments \cite{Ofner1994}. The features of the valence band of a-In$_2$O$_3$ including the main peak and the shoulders at lower energies agree very well with measurements. 

\begin{figure*}[t!]
\includegraphics[width=\linewidth]{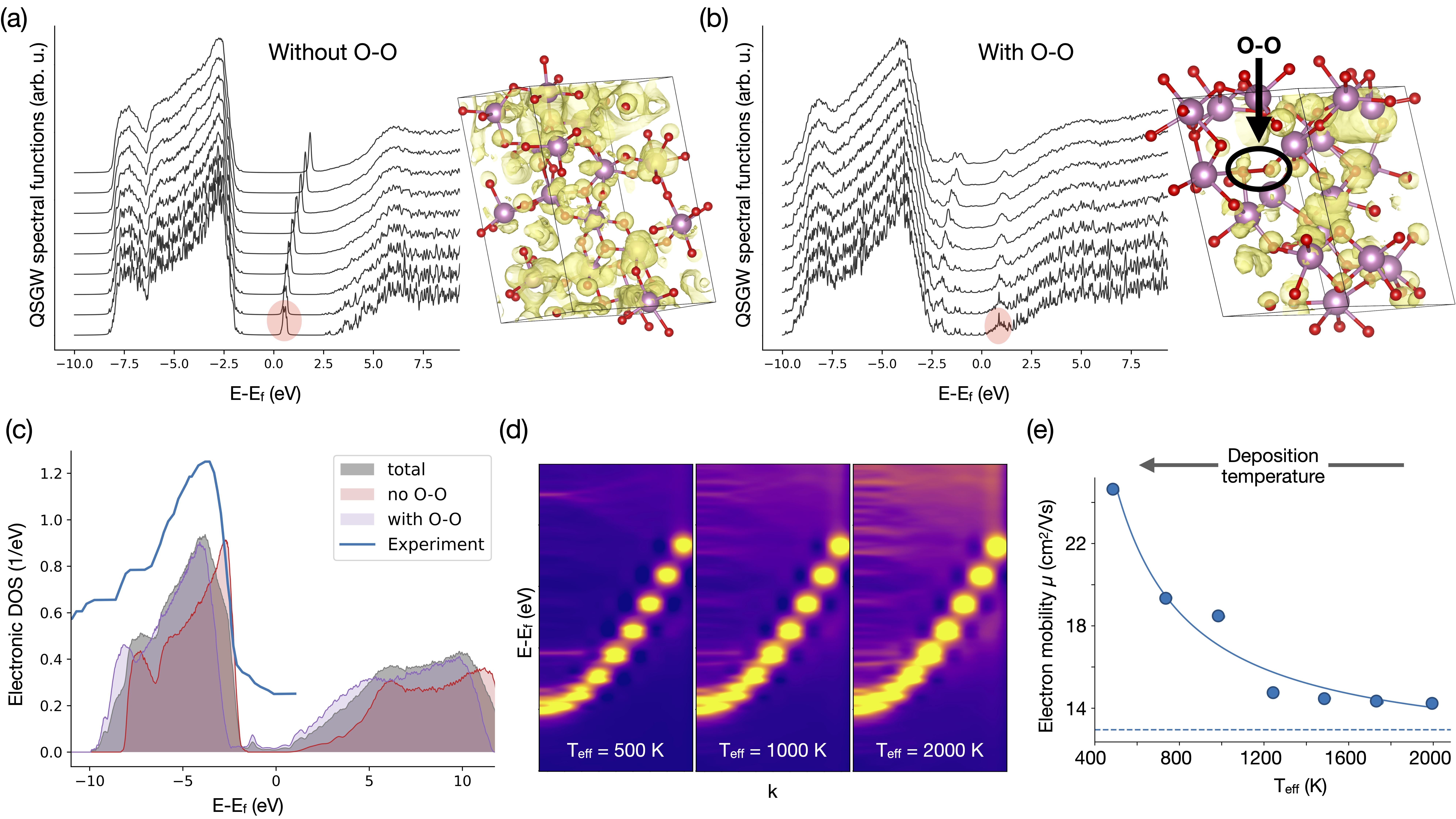}
\caption{\label{fig2} (a) The spectral functions averaged over structures with no O-O bonds and a typical CBM charge density of one of those structures. (b) The spectral functions averaged over structures that contain O-O bonds and a typical charge density for one of those structures. (c) The calculated density of states averaged over structures with and without O-O bonds compared to the total DOS and experiment. (d) Heat maps of the spectral functions for the conduction bands when the structures are ensemble averaged weighted by the Boltzmann distribution at different effective temperatures. (e) The calculated mobility of different structures as a function of the effective temperature. The correspondence with the experimental deposition temperature is indicated.}
\end{figure*} 

This model also lends itself to a procedure for calculating the effective electronic (band) structure of the amorphous phase. First, the spectral functions $A(k,E)$, measuring the probability of finding an electron with energy $E$ and wave vector $k$, are the quantities being averaged.  However, because of the random relative orientations of the local environments in the composite, $A(k,E)$ must be spherically (isotropically) averaged over all directions of ${\bf k}$ before ensemble averaging. 

Our results are presented in Fig.~\ref{fig1}(e), where isotropically and ensemble averaged $A(k,E)$ from QS\textit{GW} are shown as a function of energy for 9 different $k$-values, from $k=0$ to $k=0.16$ inverse Bohr radii ($a_0^{-1}$). The dispersion in the conduction band (red-highlighted region) is clearly visible. A more detailed view is provided in the heat maps in Fig.~\ref{fig1}(f) with the curvature of the conduction band and the energy broadening indicated. These results allow quantification of the dispersion $E(k)$ and the broadening $\Delta E$. From there, the electron effective mass $m^{*}$, relaxation time $\tau$, and the disorder-limited mobility $\mu$ are evaluated using the standard expressions:
\begin{equation}
\tau = \frac{\frac{1}{2}\hbar}{\Delta E} \, , \, \mu = \frac{|e| \, \tau}{m^{*}}.
\end{equation}
The following numerical results are obtained. The electron effective mass computed from the effective band structure is $m^{*} \approx 0.2\,m_0$, almost identical to the value for the crystalline In$_2$O$_3$, calculated using the same procedure as $m^{*}=0.22\,m_0$.  Reported experimental values fall in the range $0.18-0.33\,m_0$ \cite{Nakazawa2006,Scherer2012}. For the k-points near $\Gamma$ nearly uniform broadening of $\Delta E=0.25$ eV is found. These values result in the electron mobility of $\mu \approx 13$ cm$^2$/Vs. Finally, by taking electron concentrations in the range $10^{20}-10^{21}$ cm$^{-3}$, typically reported from transport measurements,  we estimate the disorder-limited electron mean free path to be 1-3 nm, which amounts to $\sim$5-15 In-O bond lengths.

The calculated mobility corresponds very well to the experimental mobilities measured for a-In$_2$O$_3$ deposited at low temperatures \cite{Buchholz_PNS:2013}. Generally, measured mobilities start at $\sim$10 cm$^2$/Vs for a-In$_2$O$_3$ thin films deposited at $-100$ to $-50~\degree$C, and then increase to $\sim$60 cm$^2$/Vs as the deposition temperature increases \cite{Buchholz_PNS:2013}. The reasons for this increase are discussed below.

Upon studying relationships between the electronic structure of individual random structures and their atomic arrangement, one finds that the most important feature influencing the electronic structure is whether the atomic arrangement contains O-O bonds. Of the full 1500 structures, 54$\%$ contain one or more O-O bonds (52$\%$ in the smaller sample used for QS\textit{GW} calculations). All structures without O-O defects exhibit a highly dispersive conduction band with low broadening.  If one averages the spectral functions only over those structures, no ``midgap'' states are observed, and the peaks of the spectral functions in the conduction band become sharper, as shown in Fig.~\ref{fig2}a. In contrast, if only structures with O-O bonds are considered, the dispersion breaks down and the conduction band splits into two, one set of states below the Fermi energy, and the other above, as seen in Fig.~\ref{fig2}b. This is primarily because the presence of an O-O bond changes the effective electron count in surrounding coordination polyhedra. The difference in the degree of localization of the CBM states between structures with and without O-O bonds are clearly visible from the CBM charge densities of individual structures shown in Fig.~\ref{fig2}(a) and (b) (right panels).

This different localization can also be seen in the calculated inverse participation ratios (IPR) for the states above and below the Fermi energy, shown in SI Figs.~\ref{fig:IPR} and \ref{fig:log_IPR}. For structures without O-O bonds, at energies in a range between \(\sim\)$-$2.7 and \(\sim\)$-$1.5 eV near the valence band edge, the states are localized. In contrast, all of the states in the conduction band near 1.5 eV (which lie within a very small tail of the density of states, are delocalized. Since it is expected that weak disorder can localize states near band edges~\cite{ziman_localization_1969,abou-chacra_self-consistent_1974,nichols_localization_1988}, the observed delocalization suggests that the scattering from ideal electronic states caused by  changes in coordination number without O-O bonds is small -- not large enough to appreciably localize conduction band states. For structures with an O-O bond, we observe delocalized valence band states below about $-$2.0 eV, which become localized at higher energies within the valence band, consistent with the idea that localization near band edges is preferred. In the conduction band, structures with O-O bonds show moderate localization for most states within \(\sim\)1.5 eV of the conduction band minimum, suggesting that the electronic scattering associated with the O-O defects is sufficiently strong to localize states near the band edge. 

Looking at the average densities of states in Fig~\ref{fig2}(c), calculated by considering only the structures with O-O bonds, and separately the structures without O-O bonds provides support to the idea that the electronic properties of the system best reproduce the experimentally observed properties when treated as a composite of the properties with and without O-O bonds. The density of states averaged over all 100 structures is in agreement with the experimental UPS measurements in that states at energy lower than the maximum peak are concave down, rather than linear or concave up as seen in the average over the states with and without O-O bonds, respectively. The average including states with and without O-O bonds also exhibits two large and two small shoulders at energies lower than the maximum frequency, showing much better agreement with the measured UPS spectra than the single distinct low-energy peaks observed in the densities averaging structures with and without O-O bonds separately. 

Additionally, some fluctuations in the dispersions and the CBM position relative to the common Fermi energy can be observed even in structures without the O-O bonds. They are responsible for the fine peak structure in the conduction band, visible in the Fig.~\ref{fig2}(a) (see inside red circle) and contribute to the CBM energy broadening close to the $\Gamma$ point. Our analysis indicates that they are a consequence of the fluctuations in the first-shell In-O coordination (see Fig.~\ref{fig:coord_cbm_corr}). Structures with lower fraction of 6-fold coordinated In will have somewhat higher CBM energy 
relative to those with majority of In atoms 6-fold coordinated.

Finally, both the structures containing O-O defects and those with lower fraction of the 6-fold coordinated In, are higher in energy than those without O-O and with predominantly 6-fold coordinated In, as shown in Fig.~\ref{fig:coord_cbm_corr}. Because of this, one might expect that the fraction of those structures present in the actual samples would decrease as the experimental conditions get closer to thermodynamic equilibrium. In other words, averaging over all structures corresponds to the experimental situation far away from equilibrium, that is, to the thin-films grown on cold substrates. Cold substrate conditions in the thin-film growth have previously been shown to correspond to ensemble averages at high thermodynamic temperature \cite{Caskey_JCP:2016}. The effects of increasing substrate temperatures can then be modeled by ensemble averaging properties across random structures using the Boltzmann distribution, assuming the energy of a particular configuration is given by the energy of the PBE functional. Lower effective temperature in this distribution, corresponding to higher substrate temperatures in thin film growth, will lead to the lower concentration of O-O defects and 4- and 5-fold coordinated In. In turn, this will lead to increased electron mobility by reducing the conduction band linewidth. 

The spectral functions in Fig.~\ref{fig2}(d) show a decrease in linewidth with decreasing effective (ensemble averaging) temperature. Consequently, mobility increases with decreasing effective temperature, as shown in Fig.~\ref{fig2}(e), which agrees with the experimental observation that increasing deposition temperature will lead to increased mobility for In$_2$O$_3$ thin films. The O-O defects and to some extent In-O coordination lower than 6 provide major perturbations (scattering centers) to the conduction band electronic states. Their abundance and the associated localization of conduction band states directly correlates with reduced electron mobility. 

The remaining difference between our low effective temperature electron mobility of $\sim$25 cm$^2$/Vs and the experimentally observed maximum of $\sim$60 cm$^2$/Vs in films deposited at $\sim$0$^{\circ}$C is likely related to the presence of doping (electrons), which is not accounted for in our approach. Electron doping will result in having some fraction of localized states (e.g. due to O-O defects) being occupied, effectively reducing the number of scattering channels, which will lead to the increase in the electron mobility with doping concentration. Additionally, increase in doping concentration with deposition temperature will also lead to the higher Fermi velocity and the electron mean free path. 

Increasing the deposition temperature to $\sim$25$^{\circ}$C or more leads to the appearance of crystalline features in the X-ray diffraction, which coincides with the drop in measured electron mobility from $\sim$60 back to 10 cm$^2$/Vs \cite{Buchholz_CM:2014}. Our results may offer an explanation for this drop as a consequence of the misalignment (band offset) of the CBM positions between amorphous and crystalline phases evident from Fig.~\ref{fig1}(f). The amorphous and crystalline CBM differ by $\sim$250\,meV, creating an effective barrier for electron transport. If we imagine having crystalline inclusions in an amorphous phase, measured to be in excess of 36 \% for films deposited $\sim$25$^{\circ}$C or more \cite{Buchholz_CM:2014}, their presence will lead to the drop in mobility because of the band offset. The drop will continue until the crystalline fraction is large enough to establish a percolation, at which point the mobility will begin to increase with the deposition temperature as also observed experimentally \cite{Buchholz_CM:2014}.

\section{Physical origins responsible for electron delocalization and band-like transport}
%
It is worth noting that treating amorphous phase as a collection of independent local domains and calculating an eﬀective electronic structure as an average over the domains can be justified under the following conditions. First, spatial correlations need to be relatively short-ranged so that the linear dimensions of the domains are larger than the typical correlation length. Second, no phase coherence should exist between the domains. In other words, the domain size needs to be larger than the typical electron dephasing length. Both of these conditions are likely to be fulfilled for relatively small domains (few nm) in rapidly quenched liquids (i.~e. glasses) or amorphous systems grown on cold substrates, and for temperatures around room temperature or above. The good overall agreement with the measurements indicates that the correlation length and the room temperature dephasing length in a-In$_2$O$_3$ are both shorter than the size of the cells used in our random structures.  

Next, the survival of the conduction band after both isotropic and ensemble averaging indicates that: (i) the highly dispersive conduction band is present in the large fraction ($\sim$50 \%) of the individual structures despite their low symmetry, (ii) the spatial anisotropy of the conduction band is low for all those structures, and (iii) the bonding environments in different structures are sufficiently similar to produce finite band gaps, similar number of states in the valence and conduction bands, low number of mid-gap defect states all leading to similar conduction band energies relative to E$_f$. O-O bonds are the primary defects that break these patterns in amorphous In$_2$O$_3$.
Indeed, a dispersive conduction band of the dominantly In-5\textit{s} character can be found in all structures without the O-O defects. This is true for the 100 structures calculated at the QS\textit{GW} level of theory,  and also in the remaining 1400 (at the DFT level). It is a consequence of the atomic coordination environments being similar among themselves and to those of bixbyite. Most of the In atoms are five- and six-fold coordinated by O, and the O atoms are three- and four-fold coordinated by In in the random structures. Moreover, the connectivity of In coordination polyhedra in each random structure also resembles bixbyite; most of them are edge- and corner-sharing. The existence of the dispersion then implies similar site energies and strong overlaps of the ``molecular'' orbitals (In-5\textit{s}-like) forming the conduction band. The similarity of the site energies is supported by the observation that the inverse participation ratio shows strong delocalization of these conduction states (see supplementary information Fig.~\ref{fig:IPR}). 

Additionally, the extended range and direction-agnostic \textit{s}-like character of the orbitals forming the conduction band means that connected In coordination polyhedra have substantial overlap, regardless of their orientation. This effect also contributes to the conduction band being largely isotropic in space for each individual structure, leading to the low broadening observed after the isotropic average. In contrast, the valence band states, composed primarily of less-extended and directional O-\textit{p} orbitals are significantly more localized and do not lead to any dispersive band-like features, as evident from Fig.~\ref{fig1} and the inverse participation ratio analysis in Fig.~\ref{fig:IPR}. Lastly, similarity between building blocks across all random structures implies similarity in the electronic DOS in both valence and conduction states. The consequence of this is the similar magnitude of the dispersion, and similar positions of the conduction bands relative to their respective E$_f$, all contributing to relatively low broadening of the effective conduction band. 

Hence, the existence of the dispersive and relatively sharp conduction band in more than half of random structures, and by extension in our model of the amorphous phase, is due to the strong tendency of In$_2$O$_3$ to form certain types of coordination polyhedra, their connectivity, and a low number of defects (wrong coordination environments). Similar behavior can be observed in amorphous/glassy SiO$_2$, with corner-sharing Si-O tetrahedra representing the sole building block with \textit{s}-like orbitals forming the low-energy unoccupied states giving rise to the existence of a conduction band (see supplementary materials) Fig.~\ref{fig:eff_bnd_SiO2}. The \textit{s}-like character of the conduction band minimum in both SiO$_2$ and In$_2$O$_3$ is indicated by the single degeneracy of this band.  The apparent lower dispersion of CB in SiO$_2$ is related to the near absence of edge-sharing polyhedra \cite{Wolf_JAP:2025} and the shorter radius of Si-\textit{s} orbitals. 

In conclusion, by treating amorphous phase as a composite of local environments we can quantitatively describe at the QS\textit{GW} level of theory the band like character of the 
lowest conduction band in a-In$_2$O$_3$, including both the effective mass and electron mobility. The reasons for the existence of the conduction band despite the absence of long range order include:  (i) the extended, direction-agnostic \textit{s}-like character of the orbitals forming the band, thus averaging over many neighbors; (ii)  coordination within polyhedra and their connectivity is largely preserved  in the great majority of local environments, and (iii) relatively small number of detrimental O-O defects implying a strong tendency to the formation of certain types of coordination environments. Similar behavior of amorphous SiO$_2$ (at the DFT level of theory) shown in the supplementary information (Fig.~\ref{fig:eff_bnd_SiO2} further supports this conclusion. Lastly, the fully first-principles approach presented here for describing the electronic structure of the amorphous/glassy states is general and can be applied to electronic structure properties of disordered systems more broadly.

\section{Methods}
%
\subsection{First principles random structure sampling}
A set of 1500 random In$_2$O$_3$ structures were generated with 40 atoms per unit cell in a manner that encourages numerical stability of relaxations and favors cation-anion coordination \cite{Stevanovic_PRL:2016}. For each structure, random values of lattice parameters $(a,b,c,\alpha,\beta,\gamma)$ are chosen using the constraints $0.8 * scale \leq a,b,c \leq 1.6 * scale$ and $60 \degree \leq \alpha,\beta,\gamma \leq 140\degree$. Atoms are distributed within the cell in a manner favoring cation-anion coordination and leading to homogeneous initial distribution. To achieve this, a grid is generated at the minima and maxima $(\bf{r})$ of a planewave $\cos(\bf{G}\cdot \bf{r})$, defined by reciprocal lattice vectors ${\bf G}=n_1{\bf g}_1+n_2{\bf g}_2+n_3{\bf g}_3$. Cations are placed at the maxima of this wave, while anions are placed at the minima. A gaussian is centered on each atom, and the next atom placed is placed with higher probability on a position where the sum of these gaussians is low. The scale of the structure is changed so that the minium distance between atoms is approximately 80\% of the In-O bond length in bixbyite. 

Once these structures are generated, they are relaxed using the PBE approximation in density functional theory as implemented in VASP\cite{VASP1}. Ionic relaxations are performed using the conjugate gradient algorithm so that structures relax reliably to their nearest local minimum. Pseudopotentials are constructed from the projector-augmented wave (PAW) method. The plane wave cutoff is 340\,eV, and the $\Gamma$-centered \textit{k}-point grid is generated automatically using 20 divisions of the grid. All degrees of freedom, volume, cell shape, and atomic positions are included during relaxations, and are restarted multiple times to resolve Pulay stress during relaxations. The cell shape and volume are considered converged when the hydrostatic pressure is less than 0.5 kilobar. Ionic relaxations are considered converged when the maximum force on any atom is less than 0.02 eV/\AA. 

\subsection{Quasiparticle self-consistent \textit{GW} calculations}
The electronic structure of a subset of 100 of these relaxed random structures is computed using quasiparticle self-consistent \textit{GW} (QS\textit{GW}), as implemented in the Questaal code \cite{qsgw_PRL:2004}. QS\textit{GW} is necessary here because PBE significantly underestimates the band gap of In$_2$O$_3$, so a correction to the electronic states that does not depend on the starting point, as in QS\textit{GW}, will lead to an electronic structure that can be reliably interpreted.  In QS\textit{GW}, the self-energy is quasiparticalized as described in Ref.~\cite{qsgw_PRL:2004}, that is, the self-energy is treated as a static Hermitian potential, and this is iterated to self-consistency, removing the dependence of the self energy on the starting point used for calculating the charge density. This ensures that errors in the electronic properties are systematic and well-defined, coming from specific effects not included in the random phase approximation, such as electron-hole interactions and electron-phonon coupling \cite{questaalBSE}. 
The self-energy is considered converged when the root mean squared deviation between steps is less than $4\times10^{-5}$. Electronic states around each atom are approximated using an optimized linear muffin-tin orbital basis, and higher energy states are approximated with plane waves. The cutoff radius of these different basis functions is chosen to include approximately 50 neighbors for each atom in each structure. The k-point grid for these calculations is chosen such that the number of atoms times k-point density is between 1280 and 2560, and the distance between k-point divisions along each of the reciprocal lattice vectors is approximately even. The ${\bf G}$-vector cutoff for basis envelope functions is set to 2.9 Ry$^{1/2}$, and the regular mesh for the smoothed Hankel envelope functions is 8 Ry$^{1/2}$. High-lying elements in the self-energy matrix with energy above 2.8 Ry are approximated using plane waves. Smooth Hankel energies for the basis set on In and O atoms are optimized to minimize the total energy of crystalline In$_2$O$_3$.

\subsection{Reciprocal space and real space structure functions}
The structure functions for each structure and the averages over different sets of structures were calculated as detailed in \cite{Wolf_JAP:2025}. In brief, the total X-ray structure function was computed using the Faber-Ziman formalism wherein 

$$S^X({\bf q}) = \frac{1}{\langle f(q)\rangle^2}I_a^{coh}({\bf q}) - \frac{\langle f^2(q)\rangle - \langle f(q)\rangle^2}{\langle f(q)\rangle^2}$$

Here, $\bf q$ is the scattering vector, and $q$ is its norm. $\langle f(q)\rangle=\sum_\alpha c_\alpha f_\alpha(q)$ is the average X-ray atomic form factor where $c_\alpha$ is the atomic fraction of the atoms of type $\alpha$ and $f_\alpha(q)$ is their atomic form factor. 

The pair distribution functions between given atom types are computed using

$$ g_{\alpha\beta}(r)= \langle{\frac{1}{4\pi r^2 nc_\beta}\frac{dN_{\alpha\beta}(r)}{dr}}\rangle_\alpha$$

where $dN_{\alpha\beta}(r)$ is the number of $\beta$ atoms in a spherical region with thickness $dr$ at a distance $r$ from a given $\alpha$ atom, and $\langle~\rangle_\alpha$ denotes an average over all $\alpha$ atoms. 

\subsection{Averaging of the electronic structures}
In order to average densities of states and spectral functions over random structures, the eigenvalues of the different structures must be aligned. These energies are aligned on the Fermi energy for each structure, as it represents the equilibrium concentration of electrons in a given structure. If the different structures are in contact with each other and in thermodynamic equilibrium, they should share a common Fermi energy. The Fermi energy for each structure is solved by bisection adjusting the center of a Fermi-Dirac distribution with temperature equal to 10\% of the gap between the valence and conduction bands. The Fermi energy is adjusted until the number of holes and electrons obtained from integrating over the product of the Fermi-Dirac distribution and the electronic density of states is equivalent. For states with no gap, the Fermi energy is the highest occupied state reported by Questaal. 

In order to obtain the spectral functions for each individual structure, a \textit{k}-grid consisting of concentric spheres centered on the $\Gamma$ point is used. These spheres are generated such that the linear density in the azimuthal and polar directions are equivalent, and the number of points on each sphere grows as ${\bf k}^2$ so that the surface density of \textit{k}-points is equivalent for each value of $\bf k$. The spectrum for each $k$-point is represented by a function with a Lorentzian of width 40 meV at each eigenvalue. These spectra are then averaged over all $k$ vectors in a given sphere (that share the same norm). One can then average these spectral functions over different structures aligned on their Fermi energies as discussed above. 

To average these structures at an effective temperature, the density of states and spectral functions can be weighted by a term $e^{-{\Delta E_i}/{(k_bT_\mathrm{eff}})}/Z$, where $T_
\mathrm{eff}$ is an effective temperature, $\Delta_E$ is the energy relative to the lowest energy member of the sample, and $Z=\sum_i 
e^{-{\Delta E_i}/{(k_bT_\mathrm{eff}})}$ is the partition function of the sample.

It is important to note that the procedure described here is different from the one used to calculate the effective band structure of random alloys \cite{Popescu_PRL:2010} in which the wavefunctions computed on a supercell representing an alloy are projected onto the plane waves with wavectors from the Brilloiun zone of the underlying lattice. The main difference is in the representation of the disorder as a single microstate (supercell) in the prior work on alloys and an ensemble of local microstates utilized here. In addition to these conceptual differences, the single microstate representation would still require large supercells, rendering high-accuracy QS\textit{GW} calculations challenging.

\subsection{Calculations of effective mass, linewidth, and mobility}
Several different choices can be made to characterize peaks of the spectral functions in order to calculate the effective mass and linewidth of the conduction band minimum. The simplest choice is to integrate from the Fermi energy to energy at which the integral sums to one. The expected value of the peak can be taken as $\langle E \rangle=\int_{E_f}^a A(k,E)EdE$, and the linewidth can be taken to be the standard of that distribution, $\sigma_E = \langle E\rangle^2 -\int_{E_f}^a A(k,E)E^2dE$. These functions also display an increasing background above the fermi level which can be removed, and if that background is removed, the value of $\langle E\rangle$ is better aligned with the visual peak. Because of high-frequency noise, it is not always straightforward to identify the peak as a maximum point. High frequency noise can be removed by taking a Fourier transform, removing Fourier coefficients with frequencies above a given threshold, and an inverse Fourier transform. Identifying peaks by their maxima and using their full width half max as the linewidth returns values of effective mass, mobility, and linewidth that are all close to those found through integration with and without subtracting the monotonic background.

\subsection{Acknowledgments}
%
This work is primarily supported by the National Science Foundation, Grant No. DMR-1945010 and was performed using computational resources provided by the Oak Ridge leadership Computing Facility (ALCC-ASCR award), the National Energy Research Scientific Computing Center, and the National Renewable Energy Laboratory. This work was authored in part by the National Renewable Energy Laboratory for the U.S. Department of Energy (DOE) under Contract No. DE-AC36-08GO28308. Work at the National Renewable Energy Laboratory by M.T., S.A., D.P., R.L., and M.V.S. was supported by the Computational Chemical Sciences program within the Office of Basic Energy Sciences, U.S. Department of Energy. Work in Florida (V.~D.) was supported by the NSF Grant No. DMR-2411279, and the National High Magnetic Field Laboratory through the NSF Cooperative Agreement No. DMR-2128556 and the State of Florida. Authors acknowledge the use of the National Energy Research Scientific Computing Center, under Contract No. DE-AC02-05CH11231 using NERSC award BES-ERCAP0021783 and the computational resources sponsored by the Department of Energy's Office of Energy Efficiency and Renewable Energy and located at the National Renewable Energy Laboratory. The views expressed in the article do not necessarily represent the views of the DOE or the U.S. Government. The U.S. Government retains and the publisher, by accepting the article for publication, acknowledges that the U.S. Government retains a nonexclusive, paid-up, irrevocable, worldwide license to publish or reproduce the published form of this work, or allow others to do so, for U.S. Government purposes. 


%

\newpage

\renewcommand{\thefigure}{S\arabic{figure}}
\setcounter{figure}{0}

\onecolumngrid

\begin{center}
{\large \bf Supplementary Information}
\end{center}

\begin{figure*}[h]
\includegraphics[width=\linewidth]{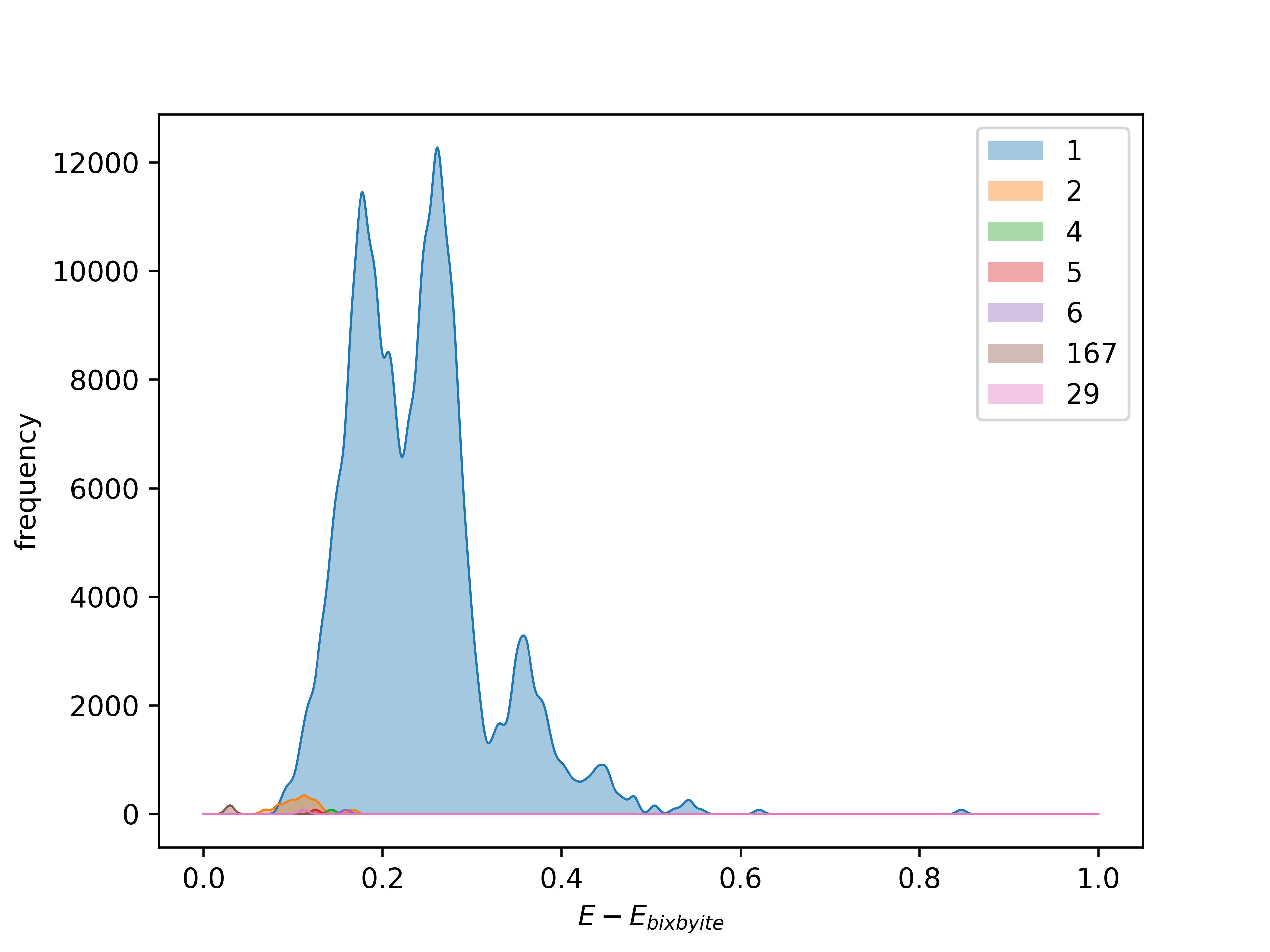}
\caption{\label{TDOS} The spacegroup-resolved thermodynamic density of states for the 1500-structure random sample of 40-atom In2O3.
}
\end{figure*} 

\begin{figure*}[h]
\includegraphics[width=0.7\linewidth]{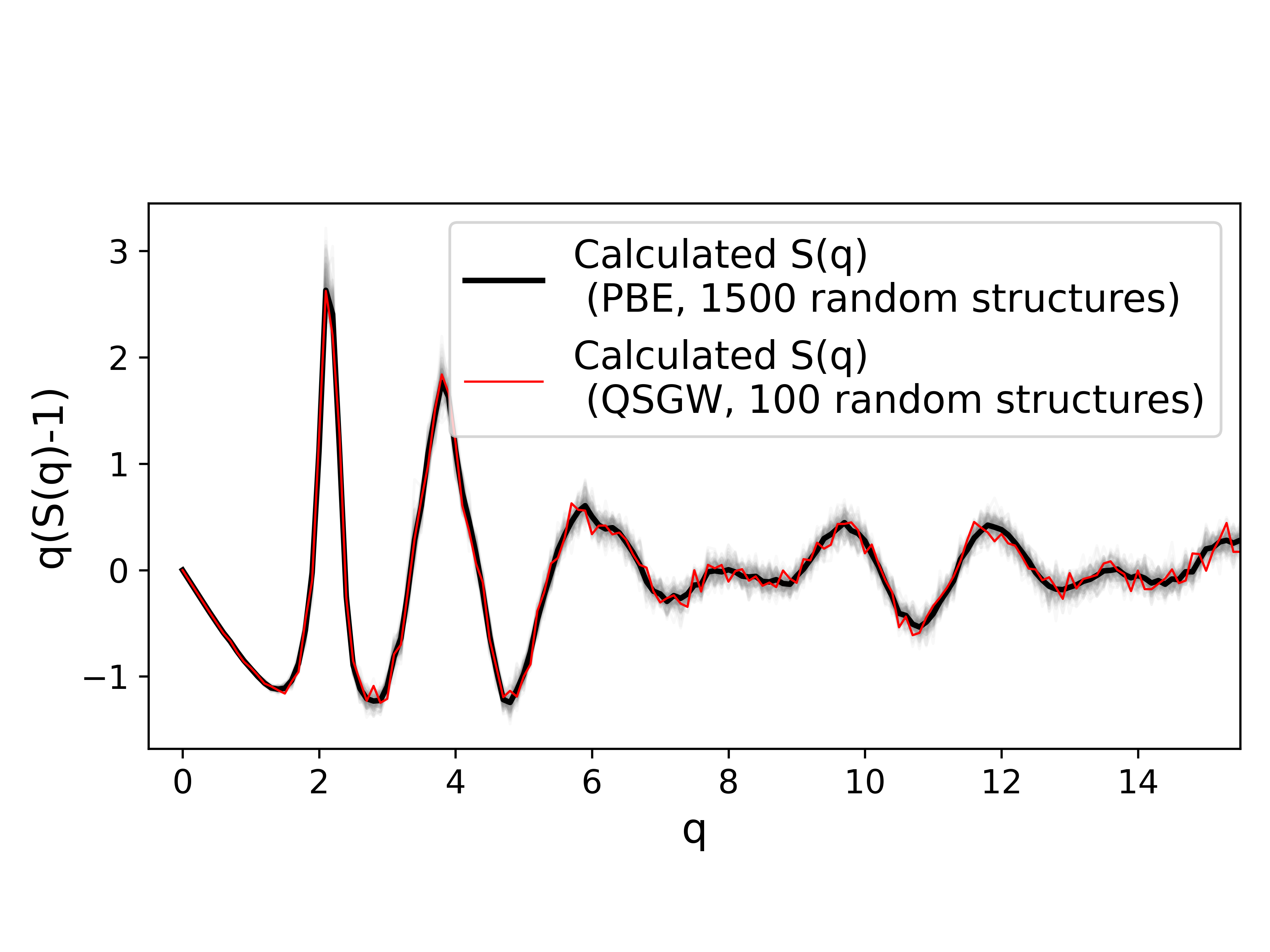}
\caption{\label{sq_conv} Comparisons of the calculated q(S(q)-1) structure factor over the set of 1500 structures (black), 100 randomly chosen sets of 100 structures (grey), and the 100 randomly chosen structures used for QSGW electronic structure calculations (red).
}
\end{figure*}

\begin{figure*}[h]
\includegraphics[width=\linewidth]{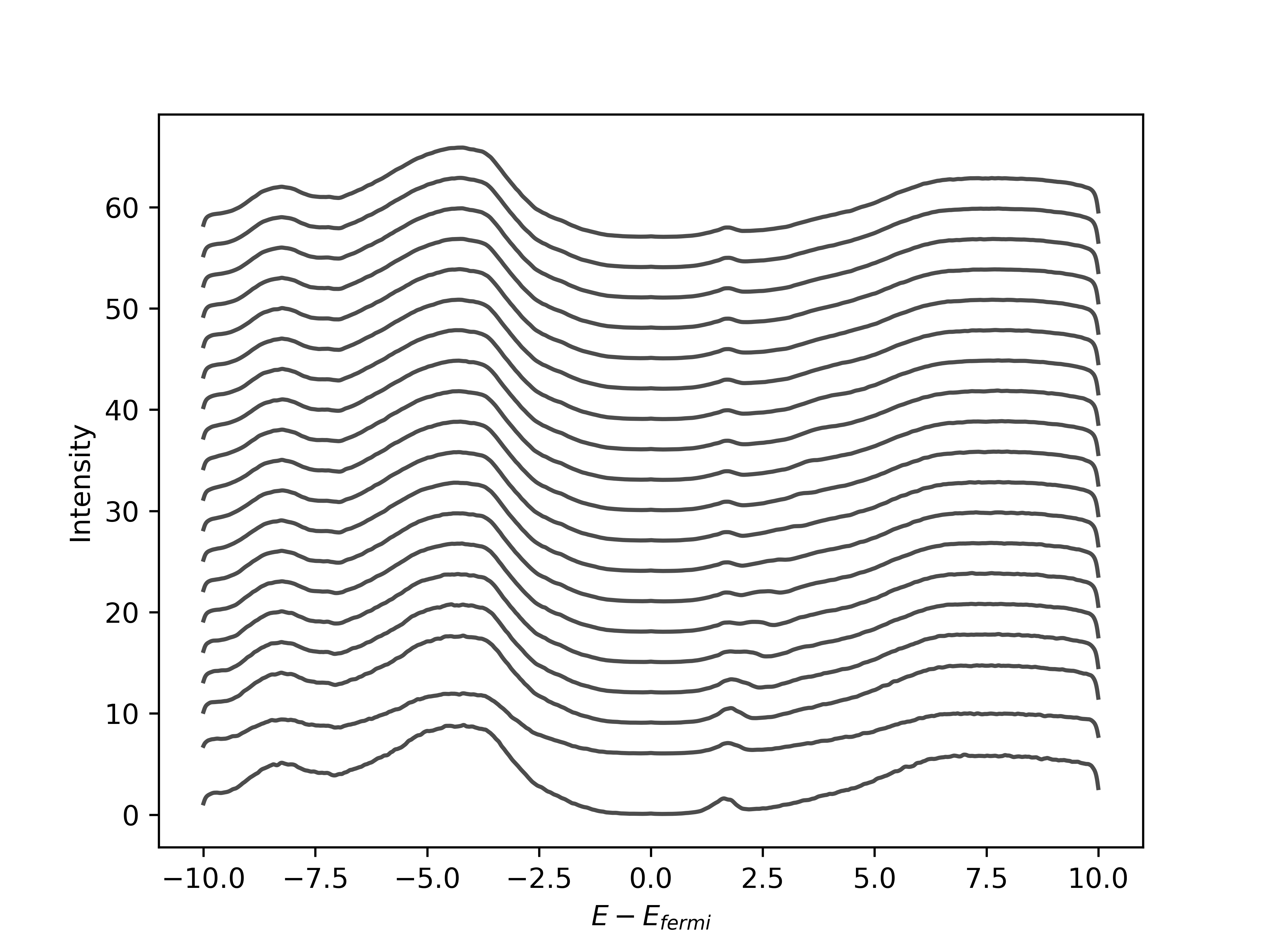}
\caption{\label{fig:eff_bnd_SiO2} The PBE spectral functions of amorphous SiO$_2$ calculated over 3000 structures, showing a flat band and a dispersive band above the Fermi energy.}
\end{figure*}

\begin{figure*}[h]
\includegraphics[width=0.7\linewidth]{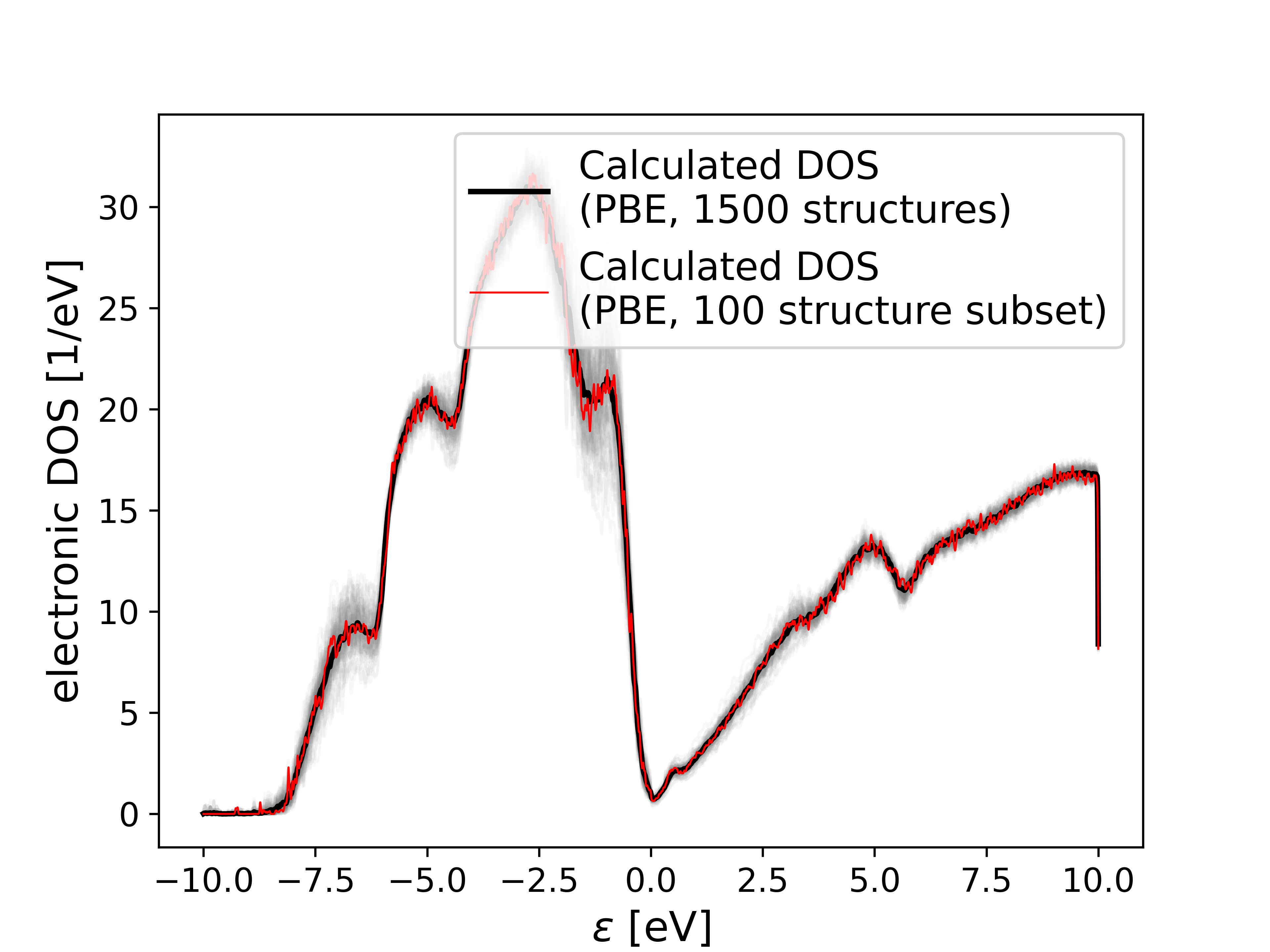}
\caption{\label{DOS_conv}
Comparisons of the calculated PBE electronic density of states over the set of 1500 structures (black), 100 randomly chosen sets of 100 structures (grey), and the density of states at the PBE level for the 100 randomly chosen structures used for QSGW electronic structure calculations (red).}
\end{figure*}

\begin{figure*}[h]
\includegraphics[width=0.7\linewidth]{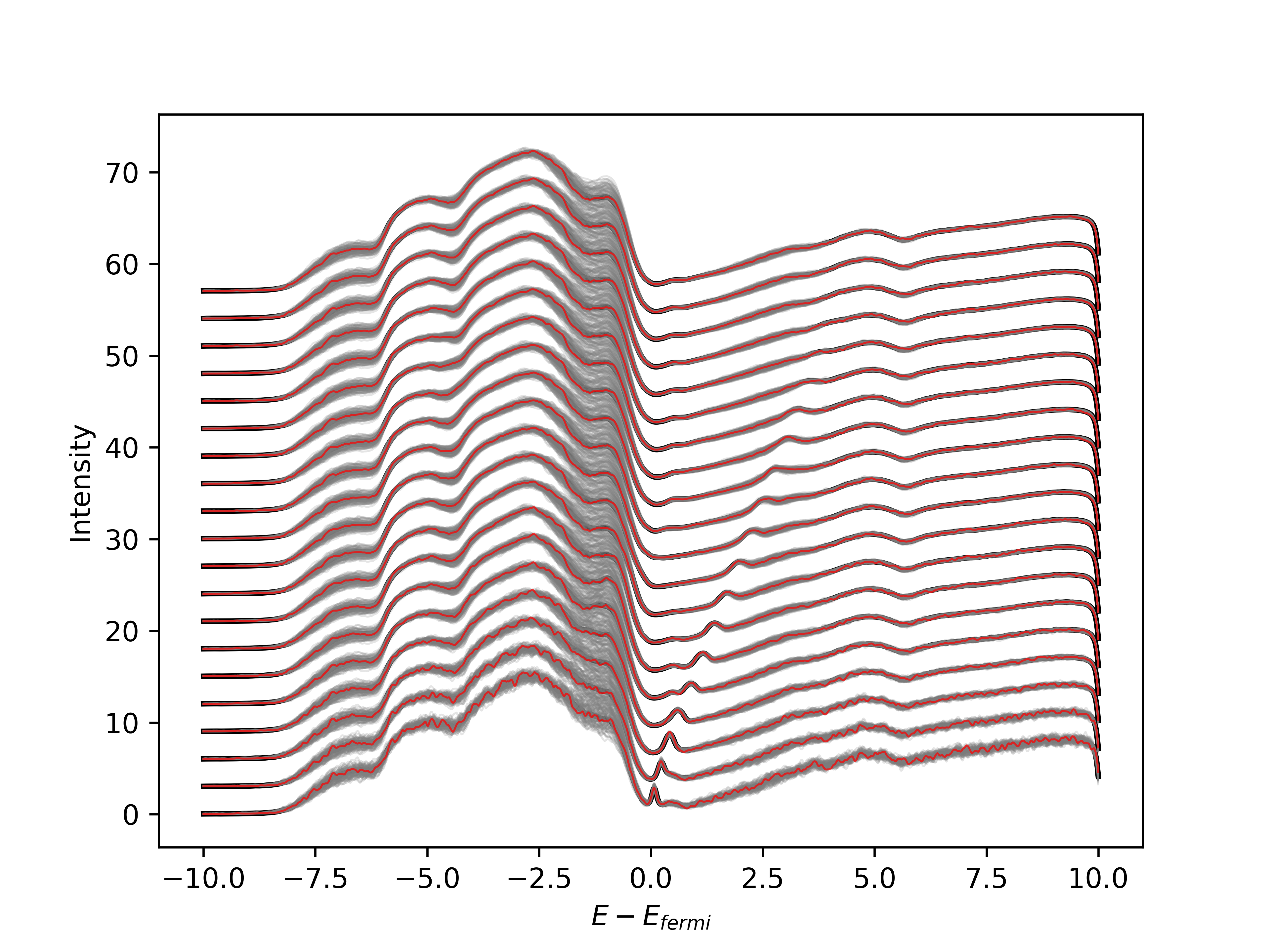}
\caption{\label{eff_bnd_conv}
Comparisons of the calculated PBE spectral functions over the set of 1500 structures (black), 100 randomly chosen sets of 100 structures (grey), and the spectral functions calculated at the PBE level for the 100 randomly chosen structures used for QSGW electronic structure calculations (red).}
\end{figure*}

\begin{figure*}[h]
\includegraphics[width=0.7\linewidth]{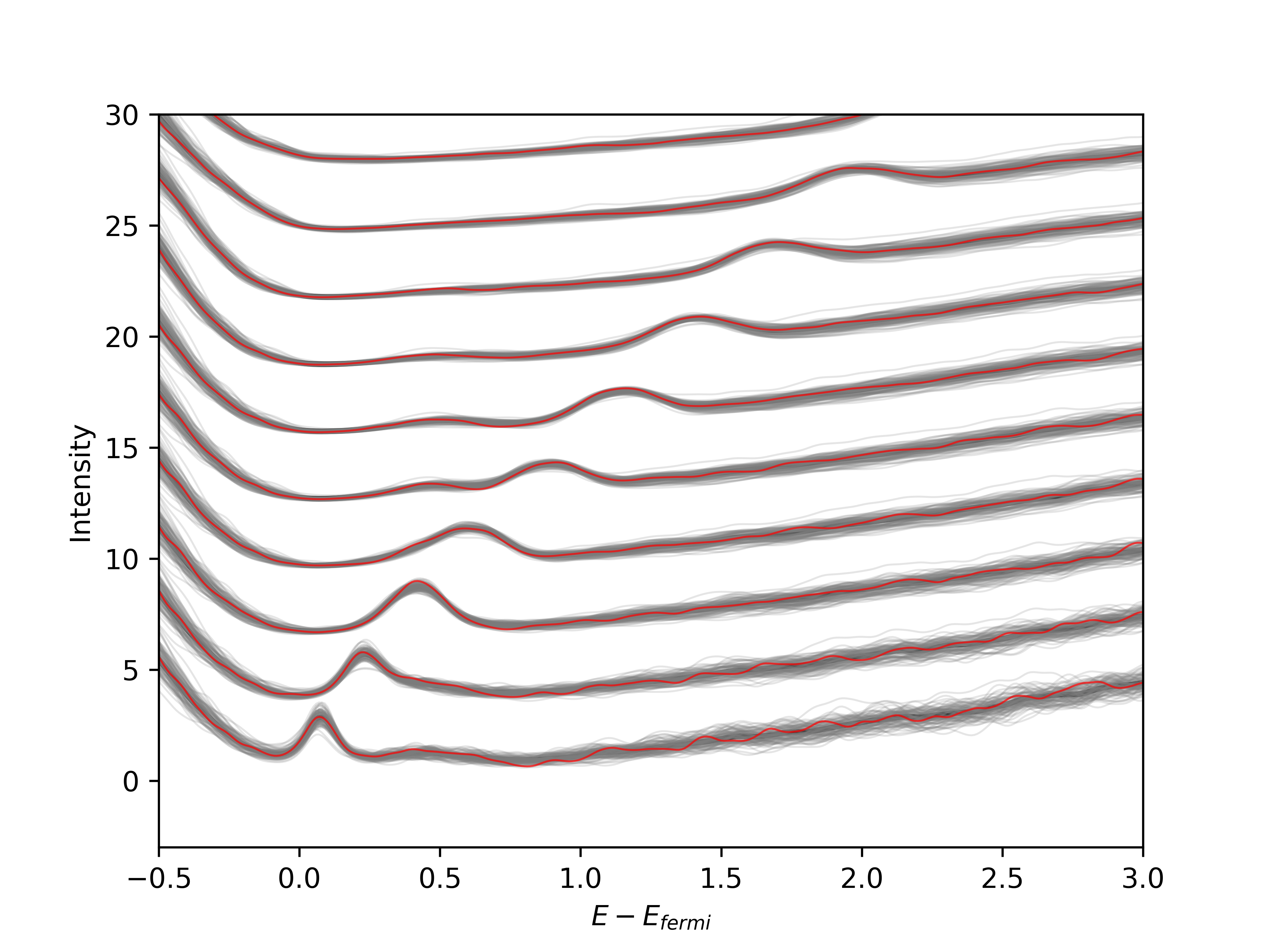}
\caption{\label{eff_bnd_conv_zoom}
Comparisons of the calculated PBE spectral functions over the set of 1500 structures (black), 100 randomly chosen sets of 100 structures (grey), and the spectral functions calculated at the PBE level for the 100 randomly chosen structures used for QSGW electronic structure calculations (red). A reduced region is shown to emphasize the resilience of the dispersive conduction band minimum feature.}
\end{figure*}

\begin{figure*}[h]
\includegraphics[width=0.7\linewidth]{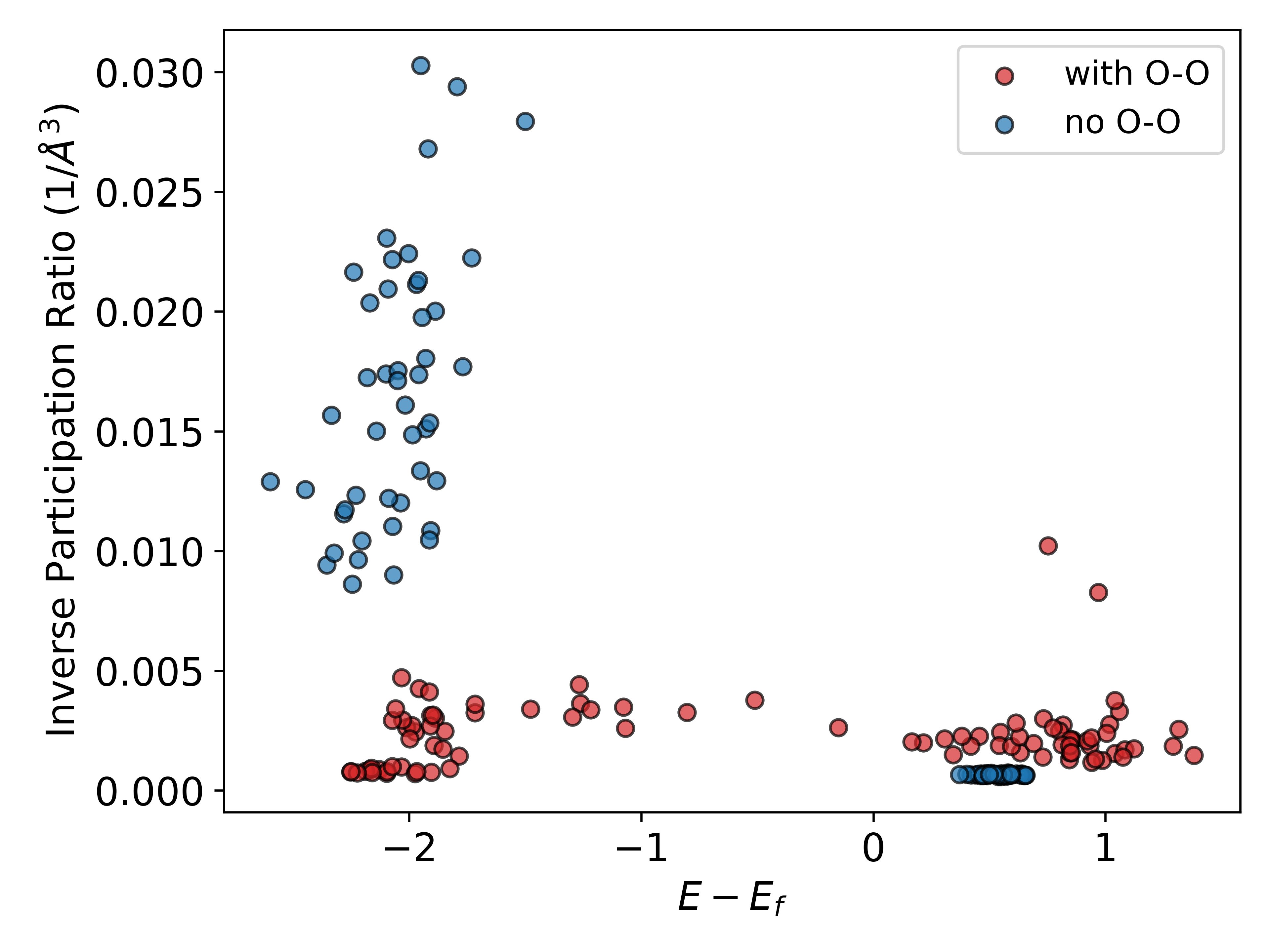}
\caption{\label{fig:IPR} Inverse participation ratio for valence and conduction states in structures with and without O-O bonds. In systems without O-O bonds, conduction states are delocalized, while valence states are localized. In systems with O-O bonds, most valence and conduction states have an intermediate degree of localization, though some valence states are more delocalized. The IPR is defined for each energy eigenstate, $\psi({\bf r})$ as $IPR = \int |\psi({\bf r}|^4 d{\bf r}/\int |\psi({\bf r})|^2 d{\bf r}$ and it has the well-known property that a state that is uniformly delocalized in a volume $V$ has $IPR = 1/V$, so  1/IPR for a delocalized state is proportional to the system volume and a localized state has a value of 1/IPR proportional to a finite volume, $V_{localized}$.}
\end{figure*}

\begin{figure*}[h]
\includegraphics[width=0.7\linewidth]{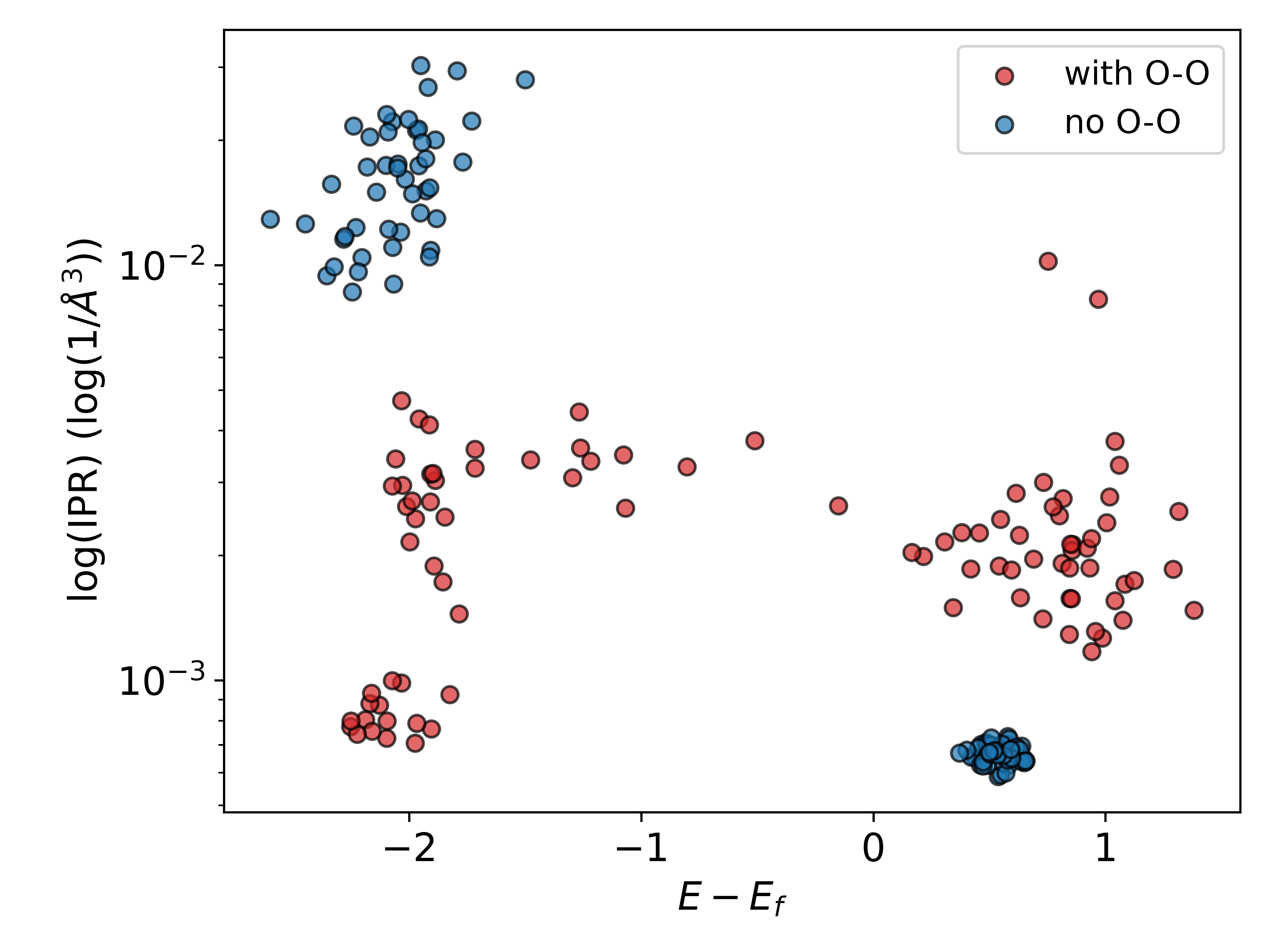}
\caption{\label{fig:log_IPR} Inverse participation ratio plotted logarithmically against the energy of the state relative to the fermi level for valence and conduction states in structures with and without O-O bonds. In systems without O-O bonds, conduction states are delocalized, while valence states are localized. In systems with O-O bonds, most valence and conduction states have an intermediate degree of localization, though some valence states are more delocalized.}
\end{figure*}

\begin{figure*}[h]
\includegraphics[width=0.7\linewidth]{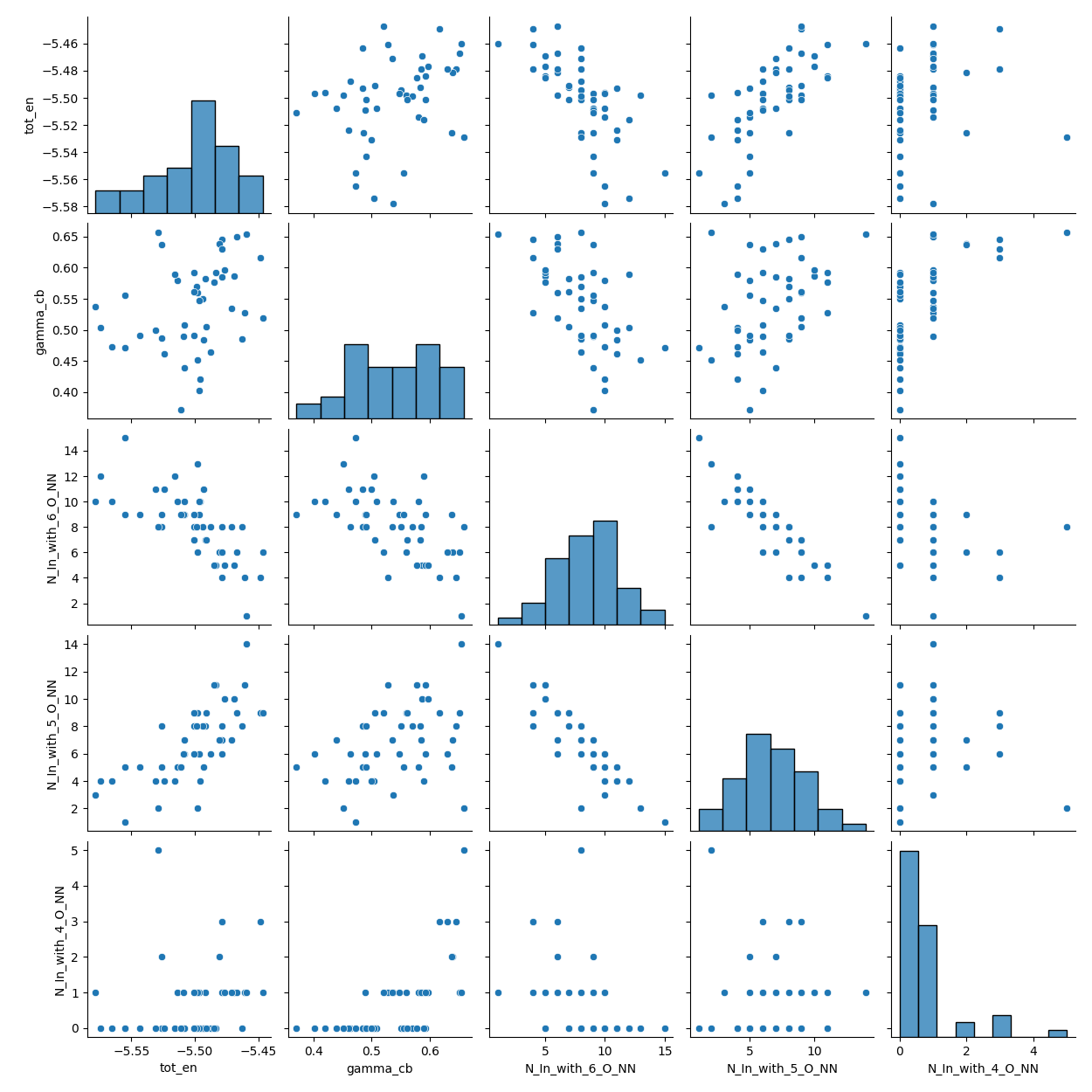}
\caption{\label{fig:coord_cbm_corr} Correlations between total energy, the position of the conduction band minimum, and the number of In with a given number of O nearest neighbors for structures in the QS\textit{GW} subset with no O-O bonds.}
\end{figure*}

\begin{figure*}[h]
\includegraphics[width=0.9\linewidth]{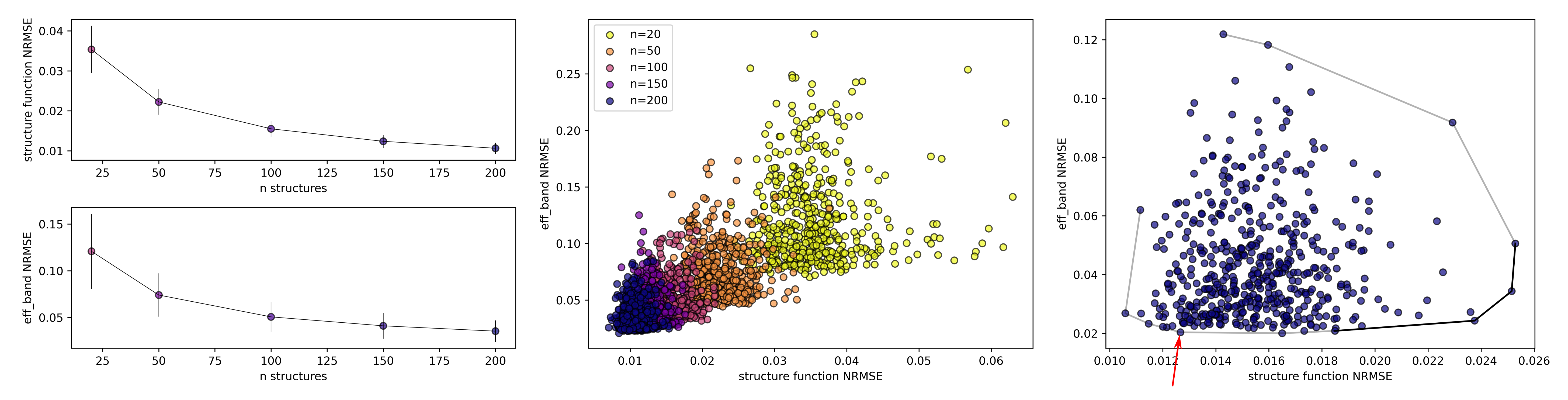}
\caption{\label{subset_choice}
a) The normalized root mean square error ($NRMSE=\sqrt{\frac{\sum_{i=1}^N(x_i-\hat{x}_i)^2}{N}}/\langle \hat{x}\rangle$) of the structure factor and effective band structure as a function of the number of structures included. b) A plot showing the distributions of NRMSE for structure factor and effective band structure. c) A convex hull for Pareto optimization to choose a set 100 of structures with low NRMSE in both structure factor and effective band structure.}
\end{figure*}

\end{document}